\documentclass[11pt,a4paper]{article}
\usepackage{jcappub}
\pdfoutput=1

\usepackage{graphicx}
\usepackage{natbib}
\usepackage{epsfig,psfig,epstopdf,float,color}
\usepackage{subfigure}
\usepackage{mybibdefs}
\usepackage{multirow}
\usepackage{rotating}
\usepackage{url}

\def\fnl{f_{\rm NL}}

\newcommand{\Po}{\mathrm{P}}
\newcommand{\A}{\mathrm{A}}

\newcommand{\dd}{\mathrm{d}}

\newcommand{\R}{\tilde{\mathrm{R}}}
\newcommand{\solM}{\mathrm{M_{\odot}}}

\title{A critical analysis of high-redshift, massive, galaxy clusters: I}

\author[a,b]{Ben Hoyle,}
\author[c]{Raul Jimenez,}
\author[c]{Licia Verde,}
\author[b]{and Shaun Hotchkiss}

\affiliation[a]{ ICC, Universitat de Barcelona (IEEC-UB), Marti i Franques 1, Barcelona
08028, Spain.}
\affiliation[b]{Helsinki Institute of Physics, P.O. Box 64, FIN-00014 University of Helsinki, Finland.}
\affiliation[c]{ICREA, Institucio Catalana de Recerca i Estudis Avancat.}

\emailAdd{benhoyle1212@icc.ub.edu}
\emailAdd{licia.verde@icc.ub.edu}
\emailAdd{raul.jimenez@icc.ub.edu}
\emailAdd{shaun.hotchkiss@helsinki.fi}

\abstract{
We critically investigate current statistical tests applied to high redshift clusters of galaxies in order to test the standard cosmological model and describe their range of validity.
We carefully compare a sample of high-redshift, massive, galaxy clusters with realistic Poisson sample simulations of the theoretical mass function, which include the effect of Eddington bias.  We compare the observations and simulations using  the following statistical tests: the distributions of ensemble and individual existence probabilities (in the $>M,>z$ sense),  the redshift distributions, and the 2d Kolmogorov-Smirnov test.
Using seemingly rare clusters from Hoyle et al. (2011), and Jee et al. (2011) and assuming the same survey geometry as in Jee et al. (2011, which is less conservative than Hoyle et al. 2011), we find that the ($>M,>z$) existence probabilities of all clusters are fully consistent with $\Lambda$CDM. However assuming the same survey geometry, we use the 2d K-S test probability to show that the observed clusters are not consistent with being the least probable clusters from simulations at $>95$\% confidence, and are also not consistent with being a random selection of clusters, which may be caused by the non-trivial selection function and survey geometry.  Tension can be removed if we examine only a X-ray selected sub sample, with simulations performed assuming a modified survey geometry.
}
\begin{document}
\maketitle

\section{Introduction}

Previously Hoyle, Jimenez, \& Verde 2011 \cite[][hereafter H11]{2010arXiv1009.3884H} claimed that the existence of a sample of spectroscopically confirmed, massive high-redshift galaxy clusters presented tension with $\Lambda$CDM at $>2\sigma$  assuming WMAP 5 priors on cosmological parameters. The authors compiled a list of prominent galaxy clusters with mostly spectroscopic redshifts greater than one and mass estimates, and conservatively chose the mass measurement for each cluster (where many existed) to reduce the tension with $\Lambda$CDM. The authors presented four possible solutions to reduce the observed tension: 1) $\sigma_8\geq0.9$; 2) All mass measurements are systematically high by $1.5\,\sigma_{\mbox{mass}}$ ; 3) a large value of primordial  non-Gaussianity characterised by $\fnl>123$; 4) uncertainties in the (non-Gaussian) theoretical cluster mass function. 

This work was independently corroborated by \cite{2010arXiv1012.2732E}, who further showed that a divergence of the mass function used in H11 appears for combinations of large masses, redshifts and values of non-Gaussianity. In response to the mass function divergence, \cite{2011PhRvD..84b3517P} constructed a non-Gaussian mass function that is stable to arbitrarily large masses and redshifts. The analysis in H11 built on earlier works which had used the existence of a single, massive high-redshift cluster to signal some tension with the current $\Lambda$CDM paradigm \cite{Jimenez:2009us,Cayon2011,Holz:2010ck}.

In all of these works, the level of tension that a cluster (or set of clusters) caused with a model, was directly associated with the probability that the cluster could have been observed in the region of mass-redshift space at greater mass and redshift ($>M,>z$) than each cluster. Throughout this work we refer to this ($>M,>z$)-method as the probability that a cluster of mass $>M$ could have been detected at a redshift $>z$, and denote this statistic by $\R$ \citep[following][]{2011arXiv1105.3630H}. 

Subsequently \cite{2010arXiv1011.0004M} introduced a new approach based on ``exclusion curves'' using the ($>M,>z$) statistic, to show how the observation of any one cluster could rule out $\Lambda$CDM at a specified confidence level and provided the community with code to calculate such exclusion curves \citep[see however][and \S\ref{mort_e} for a discussion of bias]{2011arXiv1105.3630H}. The exclusion curves have been widely adopted \citep[e.g.,][]{2011arXiv1101.1290W,Jee} but they are not ideal at identifying tension using more than one cluster.  Using the ($>M,>z$) analysis, \cite{Jee} find that many clusters are in tension with $\Lambda $CDM, three of which have $\R$ values below 0.001. Most recently using the ($>M,>z$) approach, \cite{2011arXiv1107.5617C}  showed how extreme value statistics analysis of the galaxy cluster XMMUJ0044.0-2033 \citep{santos} showed that the value of $\fnl=0$ is consistent with the data (in latest version of their paper). 

Ref. \cite{2011arXiv1105.3630H} identified that using the ($>M,>z$) statistic inevitably requires that assumptions need to be made about the selection function and survey region of a given experiment before they can be correctly interpreted. Furthermore, using a de-biased statistic and increasing the survey geometry (although this is a second order effect), he found that the tension with $\Lambda$CDM presented in H11 and \cite{2010arXiv1012.2732E} could either remain or be removed, depending on which sets of simulated clusters the observed clusters were compared with. If the observed clusters were compared with a random sample of simulated clusters, then the tension with $\Lambda$CDM remains, but if they were compared with the "Least Probable" (LP) simulated clusters (determined by clusters with the lowest $\R$ values), the tension was removed.

Given that slightly different approaches seem to yield discrepant results, here we attempt to critically examine the common measures of cluster rareness, providing comparisons with simulated data and suggest methods to account for biases.  The layout of the paper is thus: in \S\ref{currentests} we critically analyse the range of validity of current tests by applying them to simulated cluster data sets. We continue in \S\ref{theory} by introducing the theoretical mass function, and continue in \S\ref{data} by describing the data sample and discussing the effect that known and unknown selection functions have on constraining desired parameters. In \S\ref{results},  we then determine the true level of tension posed to $\Lambda$CDM from the combined set of \cite{Jee} clusters and a sample of X-ray selected clusters presented in H11, using the \cite{Jee} survey geometry assumptions, by performing a direct comparison with simulations. We then analyse a set of high redshift X-ray selected clusters with a more conservative survey geometry.  We conclude and discuss in \S\ref{conclusions}. 

Throughout the paper, unless otherwise stated,  we assume a $\Lambda$CDM model with WMAP7 \citep[][]{Komatsu:2010fb} cosmological parameters (i.e, $\Omega_{\Lambda}, \,\Omega_m, \, h, \, n_s,\, \sigma_8= 0.725,0.275, \, 0.702,\,0.968, \, 0.816$), and quote $\fnl$ using the LSS convention \citep[where. $\fnl^{CMB} \simeq \fnl^{LSS}/1.3$, see, e.g.,][]{Verde:2010wp}.

\section{How reliable are current statistical techniques?}
\label{currentests}
Recent literature has focused on two related statistical tools to ascertain if individual or ensembles of clusters are in tension with the current cosmological model.  One approach has been introduced above ($>M,>z$), and the other approach defines exclusions curves in the mass, redshift ($M,z$) plane; these curves are usually selected to trace contours of equal (expected) abundance. Here we briefly describe the tests and their range of validity and limitations.

\subsection{The ($>M,>$z) analysis.}
\label{mz}
The question:  Q1)``What is the probability $\R$, that a cluster can exist in the region ($>M\,,>z$)?'', has been used extensively in the literature as a proxy for the more interesting question: Q2) ``What level of tension with a model is caused by the existence of this cluster(s)?''.   Here we show that the answer to Q2) can only be obtained from Q1) once a careful comparison with simulations has been performed. 

The ($>M,>z$) question is asked as follows: A rectangular shaped box is placed on the ($M,z$) plane, albeit with effectively an infinite upper mass boundary, and $z$ is bounded by some upper value \citep[typically 2.2, see e.g.,][]{Jimenez:2009us,2010arXiv1009.3884H,Jee}. The theoretical mass function $n(m,z)$ is integrated over the box limits to obtain the expected abundance of clusters $A_C$ in this region, i.e.,
 \begin{eqnarray}
\label{f_nmz} A_C(M>M_C,z_{C}<z<2.2) & = & \Omega \int_{M_C}^{\infty} \int_{z_C}^{2.2} n(m,z)\dd m \dd z \;,
\end{eqnarray}  
where $\Omega$ is the survey footprint. For massive objects, the mass function decays exponentially with $z$ and $M$ and therefore the upper bounds of the box do not matter in practise.  Subsequently $\R$ is then defined as the frequency of Poisson samplings from $A_C$ which are greater than 1, divided by the total number of samplings $N$, i.e., $\R \equiv (\#P^O(A_C)\ge1) / N$.

The ($> M ,> z$) approach was used e.g., in \cite{Jimenez:2009us,Jee,Holz:2010ck,2010arXiv1012.2732E} \& H11 and in general was advocated in all the pre-2011 literature dealing with cosmological constraints from clusters. This question is well posed,  and easy to visualise, but the resulting existence probability should not be interpreted directly as the level of tension a cluster causes with a particular model \citep[as first pointed out in][]{2011arXiv1105.3630H}.

For example, suppose one were to perform many Poisson samplings from the $\Lambda$CDM mass function (aka, simulations) assuming WMAP7 priors, a survey geometry of 100 sq. deg., and a selection function of $M>10^{14}h^{-1}\solM\,,z>1.0$. After examining the distributions of  $\R$ for the simulated clusters, one finds that the ``Least Probable'' (LP) cluster from each separate simulation has a spread of ``existence probabilities" of $0.001 <\R< 0.339 $ at 95\%.

Let us now assume we have detected, followed up, and measured the mass of only one cluster $C_1$, in the above survey geometry.  We would not know if it were actually the least probable cluster in the observed region, until all other clusters had been observed and their redshifts and masses measured.

Now, if $\R$ of $C_1$ was much lower than $0.001$ we could immediately claim that this cluster is in tension with the $\Lambda$CDM model, irrespective of whether or not this cluster was the least probable in the observed region, e.g., without the identification of any other clusters. This is because no other simulated cluster (assuming WMAP7 priors) could be as rare as this cluster. We would still need to determine the level of tension, i.e. to convert the results of question Q1) to the results of question Q2),  e.g., by comparing to $\R$ from clusters simulated with different amounts of non-Gaussianity (or higher values of $\sigma_8$).

However, let us now assume that $\R$ for  $C_1$ is, for example,  $0.1$ (i.e. $>>0.001$). We cannot immediately claim that this cluster is {\bf not} in tension with $\Lambda$CDM because we will remain unsure if $C_1$ is the least probable cluster in the footprint until all others have been observed. It may turn out (e.g., if the ``true" observed universe is very different from a WMAP7 $\Lambda$CDM model) that all other clusters are more rare than $C_1$, and $C_1$ was a randomly selected cluster.

If  $C_1$ is a randomly selected cluster, i.e., from a fair sampling of the ($M,z$) distribution of all possible observable clusters, then the probability $0.1$ is in fact extremely low compared with expectations from a WMAP7 $\Lambda$CDM model. One expects randomly selected clusters (as drawn from simulations) to have existence probabilities between $0.8<\R<1$ at 95\%, assuming the above survey geometry. Until the selection function is known, we should compare two cases, the least probable, and random sets of clusters.

If we observe N clusters, $C_{1..N}$, we can compare their ensemble probabilities  with the N least probable clusters to immediately identify if tension exists.  Here we define the ensemble probability $\R_N$ as the multiplication of all individual cluster probabilities $\R$. In the case studied later in this work,  N=23 and the 95\% range of ensemble existence probabilities  (from each simulation) for the N least probable simulated clusters is $10^{-9}<\R_{23}<10^{-3}$. Does this mean that if the ensemble probability $\R_{23}$ from observed clusters were $10^{-8}$ we could conclude that there is no tension with $\Lambda$CDM? No! This claim can only be made once all the clusters in the footprint were identified and these observed clusters were then found to be the 23 least probable clusters. 

In fact, randomly selecting 23 cluster from simulations, and calculating their ensemble existence probabilities produces a range of $0.1<\R_{23}<1$ at 95\%. So if the 23 observed clusters, were consistent with being drawn from a fair sample of clusters, then their ensemble probability would be anomalously low. In \S\ref{2dkstestddes} we determine if the 23 observed clusters are consistent with being drawn from the least probable clusters, or from a fair (random) sample of clusters by comparing their distribution in ($M,z$) space and making assumptions about the survey footprint, but making no assumptions about the selection function. In \S\ref{suvgeom} we repeat the analysis but make assumptions  about the survey geometry motivated by the data.

\subsubsection{Why ($>M,>z$)?}
To highlight the arbitrariness of calling the ($>M,>z$) method an existence probability, consider this natural extension. Why should we restrict ourselves to the easily calculated, but arbitrary,  ($>M,>z$) boundary contours, e.g, what dictates that the box should be placed at right angles to the ($M,z$) axis, and not at an incline of $X\%$, or have curved instead of straight boundaries?  One could simply squash or rotate the $>M,>z$ box by $X\%$ and obtain a new existence probability $\R^*$ which would be equally as justified as the original existence probability $\R$.  Only when the value of $\R$ (or  $\R^*$) is compared with a similar analysis on simulations can it be converted to a level of tension with a model. The $>M$ boundary (and the $>z$ boundary), and the shape of the box is therefore arbitrary.

While less motivated physically, one could then be equally justified to place any shaped closed contour on the ($M,z$) plane and calculate a statistic.  However to interpret the result of this statistic in terms of tension with model, one needs to perform similar comparisons with simulations. 

$\;$\\
We conclude that the quantity $\R$ obtained from any such ($>M,>z$) analysis, although easily calculated, is not equivalent to the level of tension present with a model.  The obtained value of $\R$ must be carefully compared with a similar analysis performed on simulated clusters before it is converted into a level of tension that a cluster, or set of clusters, causes with a model. See \citep{2011arXiv1105.3630H} for further discussion of this issue.

\subsection{Single cluster exclusion curves.}
\label{mort_e} 
In this approach the ``rareness statistic" is defined by the region above a line of equal $\R$ in the ($M,z$) plane \citep{2010arXiv1011.0004M}. As with the ($>M,>z$) analysis, to determine how rare any observed cluster is, one must first assume some form for the selection function and survey geometry to identify which region of the ($M,z$) plane has been observed. Let us for now assume that a wide region of ($M,z$)  has been observed. One should then Poisson sample from the mass function many times (additionally varying model parameters, if desired) and determine a line, which again is arbitrarily defined \citep[see below, and][]{2010arXiv1011.0004M} above which only $N\%$ of clusters sit. This line is dependent on the full geometry of the survey, i.e., redshift range, footprint, selection function.

\begin{figure}
   \centering
\includegraphics[scale=0.4, clip=true, trim=20 15 15 35]{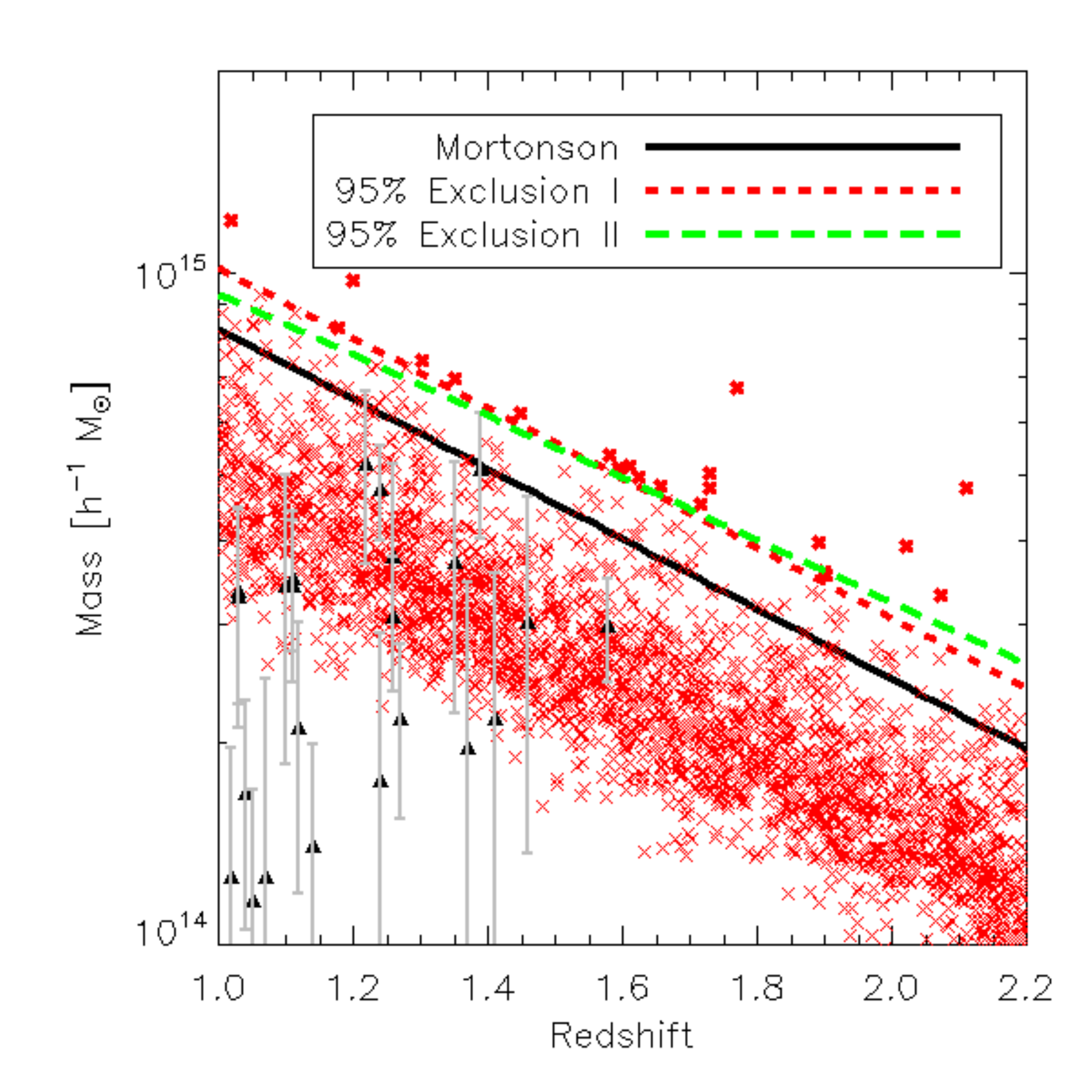}
   \caption{   \label{MortSim}  The distribution of clusters in the ($M,z$) plane obtained by Poisson sampling from the theoretical mass function assuming WMAP7 priors on cosmological parameters. We show the 5 rarest clusters from each of the $425$ simulations described in the text, by the red crosses. For reference, we also show the 23 observed clusters by the black triangles. We over plot the Mortonson et al. 95$\%$ exclusion curve for 100 sq. deg., a rescaled Mortonson et al. curve (red dashed line) which is a true 95$\%$ exclusion curve, and a different, arbitrary,  95$\%$ exclusion curve (green dashed line).}
\end{figure}

In Fig. \ref{MortSim},  we demonstrate this rareness statistic by plotting 5 rare (actually the 5 least probable, defined in the $\R$ sense) clusters from each of the $425$ Poisson simulations of the mass function  (described in \S \ref{sim}) in the ($M,z$) plane by red crosses. For reference we also show the distribution of the 23 observed clusters by the black triangles. We additionally show the popular Mortonson et al.  (Ref. \cite{2010arXiv1011.0004M}) 95$\%$  exclusion curve (black solid line) and {\it two} genuine, (but poorly motivated in the sense of \S\ref{mz}) 95$\%$ exclusion curves (green and red dashed lines). We remind the reader that such exclusion curves are only applicable in the case of testing the tension with a model caused by one cluster, and are completely insensitive to the constraining power obtained from observing multiple clusters. We note that \cite{2010arXiv1011.0004M} do present a prescription for calculating the tension caused by N clusters, by encompassing both the lowest mass and the lowest redshift clusters, but the constraining power is not optimal.

We find that 110 simulated clusters (note that these are not the re-sampled simulations which account for the Eddington bias, see \S\ref{sim}) sit above the Mortonson et al. 95$\%$ exclusion curve (black solid line), whereas one would expect only 21 (i.e., 5\% of the 425 simulations) clusters. The green dotted line, is a vertical rescaling of the Mortonson line by 23$\%$ and correctly excludes $95\%$ of the simulated clusters. As an example of the arbitrariness of lines of constant $\R$, we also show another, but less easily calculated exclusion curve by the red dashed line, which is a rescaling and rotation of the Mortonson exclusion curve, and also excludes 95$\%$ of the simulated clusters, but a different 95$\%$. Note how some clusters under one metric (specified by the green dashed line) are unlikely, but are not unlikely under a different metric (specified by the red dashed line). 

Therefore, to compute true 95$\%$ exclusion curves one should first include the survey geometry, and simultaneously carefully specify an unbiased rareness metric \citep[see][]{2011arXiv1105.3630H}, or a metric calibrated to simulations (as here).

While one may assume some metric and infer that a cluster is in disagreement with a model at some specified confidence level, and be correct in doing so, this statement would however not be invariant for all metrics. This leads us to wonder if no observation of any one exceptional cluster can consistently, to every metric, rule in, or out a model. One could imagine that we should instead compute the line which contains 100$\%$ of the data, but this is impossible to construct because repeated Poisson sampling a number close to, but greater than 0, as in the case for the abundance value of any point in the ($M,z$) plane, will eventually return a number $\ge 1$.

As stressed by \cite{2011arXiv1105.3630H}, the ($>M,>z$) statistic considers a much smaller region on the $(M,z)$ plane than the exclusion curve passing through the minimum mass and redshift of the ($>M,>z$). But, in addition, once an exclusion curve (say 95\%) has been drawn, the rareness statistic does not care if the  5\% of expected objects to be above the line are just above the line  (as expected in the standard paradigm) or far in the region of high mass and redshift (which would signify tension).  This rareness statistics would not claim any tension in either case.  The ($>M,>z$) approach on the other hand, as it is applied using the $M$ and $z$ of all observed clusters,  would detect tension if one of the observed clusters were at really high masses and/or redshifts, once it were compared with simulations.

$\,$\\
The rest of this paper is devoted to the use of robust statistical techniques in order to describe the amount of tension a set of clusters provides with a model, and to compare models and fit parameters. In particular we calibrate the performance and possible biases of these techniques with simulated clusters samples drawn from known distributions.  

\section{The (non-)Gaussian cluster mass function}
\label{theory}
The mass function describes the number of clusters $n$, per unit mass, per unit redshift, and can be written in the form
\begin{eqnarray}
\label{f_nmz} n(M,z) & = &  \frac{\bar{\rho}}{M} f \Big(-\frac{\dd \ln \sigma_M}{\dd \ln M} \Big) \;, 
\end{eqnarray}  
where $\sigma_M$ is the {\it rms} variation of the density field smoothed on scales $M$, and $\bar{\rho}$ is the background matter density. We use the icosmo\footnote{http://www.icosmo.org/} package \citep{Refregier:2008fn} to calculate $\sigma_M(z)$, co-moving distances and other cosmology-dependent parameters. We use the functional form of $f$ \citep[see Eq. 3 of][]{Tinker:2008ff} given by
\begin{eqnarray}
\label{f_b4} f & = & \A  \Big[ \big(\frac{\sigma}{b} \big)^{-a}  +1 \Big] e^{-c/\sigma^2} \;,
\end{eqnarray}  
with $A = 0.186(1 + z)^{-0.14}$, $a = 1.47(1 + z)^{-0.06}$, $b = 2.57(1 + z)^{-0.011}$ and $c = 1.19$ as adjusted to fit large N-body simulations \citep{Tinker:2008ff}, which is suitable to describe the abundance of clusters with a mass defined within the radius at which the average enclosed density contrast is $200\rho_m$, i.e. $M_{200}$.

When using the non-Gaussian (described by $\fnl$) corrections to the theoretical mass function, we employ the \cite[][hereafter MVJ]{Matarrese:2000iz} formalism \citep[see also,][]{LoVerde:2007ri, Maggiore:2009rx, D'Amico:2010ta,2011PhRvD..84b3517P}. The non-Gaussian corrections are typically written as the ratio of the non-Gaussian to Gaussian mass functions ${\cal R}$, 
\begin{eqnarray}
{\cal R}(\fnl,M,z) &=& \frac{n(M,z,\fnl)}{n(M,z,\fnl=0)} \; ,
\end{eqnarray}
and can be written as
\begin{eqnarray}
\label{eq:ratioMVJellips}
&&{\cal R}_{NG}(M,z,f_{NL})=  \exp\left[\delta_{ec}^3
\frac{S_{3,M}}{6 \sigma_M^2}\right] \times \\
& &\!\!\!\! \left| \frac{1}{6}
\frac{\delta_{ec}}{\sqrt{1-\frac{\delta_{ec}S_{3,M}}{3}}} 
\frac{dS_{3,M}}{d\ln \sigma_{M}}  %\nonumber \\
\!+\! \sqrt{1-\frac{\delta_{ec} S_{3,M}}{3}}\right| \; ,
\nonumber 
\end{eqnarray}
where $\delta_{ec}$ is the critical density for ellipsoidal gravitational collapse, whose value is fixed to fit simulations \cite{Wagner:2010me}  \citep[although see][]{2011PhRvD..83b3521D}  and $S_{3,M}=S_{3,M}(\fnl)$ describes the (non-Gaussian) skewness of the initial density field.  The MVJ prescription has recently been shown \citep[][Wagner et al. 2011 in prep.]{Wagner:2010me} to be a good fit to N-body simulations for $\fnl \leq 500$ in the regimes; $z\leq 1.0$ \& M$\leq 10^{15} M_{\odot}/h$, $z\leq 1.5$ \& M$\leq 5\times10^{14} M_{\odot}/h$, and $z\leq 2.0$ \& M$\leq 1\times10^{14} M_{\odot}/h$.

\section{Data}
\label{data}
\begin{center}
\begin{table*}
  \begin{tabular}{r r r r r r} 
Cluster Name &Redshift  &  M$_{200}$ $10^{14}  \solM$ & Method & $\R$ & Mass reference \\ \hline
RCS0221-0321  & $1.02$ & $1.80^{+1.30}_{-0.70}$ & WL & $0.992$ &\cite{Jee}\\
WARPSJ1415+3612 $^X$  & $1.03$ & $4.70^{+2.00}_{-1.40}$ & WL & $0.706$ &\cite{Jee}\\
RCS0220-0333  & $1.03$ & $4.80^{+1.80}_{-1.30}$ & WL & $0.709$ &\cite{Jee}\\
RCS2345-3632  & $1.04$ & $2.40^{+1.10}_{-0.70}$ & WL & $0.989$ &\cite{Jee}\\
XLSSJ022403.9-041328* $^X$  & $1.05$ & $1.66^{+1.15}_{-0.38}$ & X-ray & $0.997$ &\cite{stott}\\
RCS2156-0448  & $1.07$ & $1.80^{+2.50}_{-1.00}$ & WL &  $0.916$ &\cite{Jee}\\
RCS0337-2844  & $1.10$ & $4.90^{+2.80}_{-1.70}$ & WL &  $0.567$ &\cite{Jee}\\
RDCSJ0910+5422 $^X$  & $1.11$ & $5.00^{+1.20}_{-1.00}$ & WL &  $0.595$ &\cite{Jee}\\
ISCSJ1432+3332  & $1.11$ & $4.90^{+1.60}_{-1.20}$ & WL & $0.603$ &\cite{Jee}\\
XMMUJ2205-0159 $^X$  & $1.12$ & $3.00^{+1.60}_{-1.00}$ & WL & $0.888$ &\cite{Jee}\\
RXJ1053.7+5735(West) $^X$  & $1.14$ & $2.00^{+1.00}_{-0.69}$ & X-ray & $0.989$ &\cite{stott}\\
XLSSJ0223-0436 $^X$  & $1.22$ & $7.40^{+2.50}_{-1.80}$ & WL & $0.119$ &\cite{Jee}\\
RDCSJ1252-2927  $^X$ & $1.24$ & $6.80^{+1.20}_{-1.00}$ & WL &  $0.094$ &\cite{Jee}\\
ISCSJ1434+3427  & $1.24$ & $2.50^{+2.20}_{-1.10}$ & WL &  $0.806$ &\cite{Jee}\\
ISCSJ1429+3437  & $1.26$ & $5.40^{+2.40}_{-1.60}$ & WL & $0.327$ &\cite{Jee}\\
RDCSJ0849+4452 $^X$  & $1.26$ & $4.40^{+1.10}_{-0.90}$ & WL & $0.517$ &\cite{Jee}\\
RDCSJ0848+4453 $^X$  & $1.27$ & $3.10^{+1.00}_{-0.80}$ & WL &  $0.839$ &\cite{Jee}\\
ISCSJ1432+3436  & $1.35$ & $5.30^{+2.60}_{-1.70}$ & WL & $0.265$ &\cite{Jee}\\
ISCSJ1434+3519  & $1.37$ & $2.80^{+2.90}_{-1.40}$ & WL &  $0.636$ &\cite{Jee}\\
XMMUJ2235-2557 $^X$  & $1.39$ & $7.30^{+1.70}_{-1.40}$ & WL & $0.035$ &\cite{Jee}\\
ISCSJ1438+3414  & $1.41$ & $3.10^{+2.60}_{-1.40}$ & WL &  $0.584$ &\cite{Jee}\\
XMMXCSJ2215-1738 $^X$  & $1.46$ & $4.30^{+3.00}_{-1.70}$ & WL &  $0.335$ &\cite{Jee}\\
XMMUJ0044.0-2033** $^X$  & $1.57$ & $4.25^{+0.75}_{-0.75}$ & X-ray &  $0.152$ &\cite{santos}\\
\hline
  \end{tabular}
    \caption{  \label{highzclustable} The high-redshift massive cluster sample. We show the cluster name, redshift, the mass and $1\,\sigma$ errors, an indicator of the mass measurement technique (e.g,. weak lensing as WL), the $>M,>z$  probability $\R$, and the mass reference. * indicates a crude conversion from  $M_{500}$ to  $M_{200}$ assuming an NFW profile. ** indicates the mass has been estimated and the error made symmetrical, as derived from the values quoted in \cite{santos}. The $X$ superscript denotes X-ray selected clusters.}
\end{table*}
\end{center}
The high-redshift cluster data set is drawn from the literature after imposing strict redshift ($z>1.0$) and mass ($M_{200\rho_m}>10^{14}\solM$) limits, and comprises 23 galaxy clusters. We imposed the high mass and redshift cuts because the non-Gaussian mass function is most sensitive to these clusters \citep[see, e.g., Fig 7 of reference][]{Matarrese:2000iz}. In addition, clusters of masses lower than our selected cut would be missed by most observations. Of course this is not a complete sample: there may be many more clusters above the mass and redshift cut that have not been observed. Furthermore the selection function is not known and likely extremely complex, see \S \ref{selfn}. 

Table \ref{highzclustable} presents the cluster's name, the spectroscopic redshift, the cluster mass and mass error in units of $10^{14} \solM$ (analysed assuming $h=0.7$)  in terms of $M_{200\rho_m}$, the ($>M,>z$) existence probability $\R$ and the reference to the mass measurement. In essence, this table is partially a combination of Table 1 of H11 and Table 2 of \cite{2010arXiv1009.3884H} but whereas in H11 we conservatively chose each cluster's mass estimate which caused the least tension with $\Lambda$CDM, here we use the most robust (e.g. weak lensing, or X-ray) mass estimates. 

The mass of cluster XLSSJ022403.9-041328 is quoted in terms of $M_{500}$ \citep{0709.2300}, which we converted to $M_{200}$ using a naive NFW profile \cite{1996ApJ...462..563N}, but note that this cluster does not provide significant constraining power. The cluster XMMUJ0044.0-2033 has a value of $M_{200}$ quoted in the range $3.5-5\,\times10^{14}\solM$ \citep{santos}, for which we assumed a central value and an error of $4.25\pm0.75\,\times10^{14}\solM$. 

Many of the clusters in Table \ref{highzclustable}  were compiled from Super Nova searches \citep{2005hst..prop10496P}, and from other high redshift (e.g. optical) cluster surveys \citep[e.g.,][]{2005ApJS..157....1G},  and from X-ray surveys, which we highlight by the superscript $X$.  The heterogeneous selection of these clusters makes the full modelling of  the survey geometries and selection function non trivial. To compare our analysis with the analysis of H11 and \cite{Jee}, we choose to use the  survey geometry presented in \cite{Jee}, namely a redshift range of $1<z<2.2$ and a 100 sq. deg. footprint. We compare the observed clusters with all simulated clusters above $10^{14}\solM$. This survey geometry is less conservative than that present in H11.

\subsection{ X-ray selected clusters}
In section  \S\ref{suvgeom} we analyse the sub sample of 12 X-ray selected clusters, with sets of 12 simulated clusters assuming a modified, survey geometry motivated by the following argument. The combined high redshift X-ray survey geometry is composed of many individual flux limited X-ray surveys, each with differing observation depths and with partially (or fully) overlapping footprints \citep[e.g.,][]{XCS,Finoguenov:2009mf}. Thus the combined X-ray survey geometry has a variable flux limit across the footprint making it difficult to model. Moreover  galaxy and cluster redshifts become progressively more difficult to obtain at higher redshift.  For this conservative analysis using only the 12 X-ray selected clusters in \S\ref{suvgeom}, we restrict the redshift range of the survey geometry to be $1<z<1.6$, and choose a 200 sq. deg. survey footprint.

\subsection{The selection function}
\label{selfn}
The selection function describes the completeness and purity of the cluster sample.  Without full knowledge of the selection function one is unable to ascertain if all  clusters in a particular region of mass and redshift space have been observed, and thus only the presence of clusters, and not their absence, can be used for parameter estimation.

For example, X-ray surveys are limited by flux, not mass. This is important because a cluster may be detected in a flux limited survey because it hosts an active galactic nuclei, or is undergoing a merger, and is then followed up to obtain a redshift.  A subsequent weak lensing (hereafter WL) mass measurement of this cluster, or a higher resolution exposure, would reveal it to be less massive than expected, yet it would still have made the selection criteria used in this work because of the mass measurement and redshift. The reverse is also true, imagine a mechanism which over cools the gas of massive clusters dropping a potentially massive, high redshift cluster out of the sample. Additionally not all potentially high redshift clusters have been followed up to obtain a redshift, and thus would not pass our inclusion criteria.

Thus in this analysis, we cannot currently claim that we have detected all the clusters within a particular region of the mass and redshift plane. We highlight this level of incompleteness by comparing the expected number of clusters ($\sim613$ above $10^{14}\solM$ and between $1.0<z<2.2$ assuming the Gaussian-mass function in \S\ref{theory} and WMAP7 cosmology and a footprint of 100 sq. deg.) with the total number (23) of clusters in Table \ref{highzclustable}.

%253276 -- full sky
% 613 in 100 sq. deg.

Even in this pathological case, parameters or models can still be favoured and bounded from one direction, even if the final $\chi^2$ value may be arbitrarily high. For example, if one knows the survey footprint and identifies a group of high redshift massive clusters, but  is unsure of the selection function, one may still infer that $\fnl>X$ (or $\sigma_8>Y$) is a better fit to the data than $\fnl<X$ (or $\sigma_8<Y$), as future cluster detections within the same survey footprint (e.g., from subsequent followup) would confirm, or boost the previously inferred values. Once the selection function is known, e.g., by follow up of all possible clusters, both the presence and absence of clusters could be used to bound parameters from above and below \citep[e.g., as performed by][]{2011arXiv1101.1290W}.  In a follow up paper (Hoyle et al. in prep.) we address the problem of obtaining constraints on parameters using data drawn from a  survey with an unknown selection function.  However, for now, we  proceed by assuming knowledge of the selection function over the survey geometry.  We calibrate our methods by direct comparisons with simulated data. 

We cannot claim that the clusters in Table \ref{highzclustable} are the rarest clusters in the survey footprints, until all other clusters have been observed. We do however, not expect them to be drawn from a purely random distribution, as more massive clusters have typically higher fluxes, and are thus easier to detect.

\section{Standard statistical tests and results}
\label{results}
Assuming the survey geometry presented in \cite{Jee}, namely, a redshift range of $1<z<2.2$ and a footprint of 100 sq. deg., we show that the low values of $\R$ for individual clusters, and the combined values of $\R_{23}$ are compatible with values of  $\R_{23}$ for the LP simulated clusters, but they are anomalously low when compared with a randomly selected sample of simulated clusters. Recall that the $\R$ statistic encodes the redshift and mass information of the cluster into one number.  For this reason we further present redshift distributions of the observed and simulated clusters which suggests that the observed clusters are not compatible with being the LP clusters from simulations. We then formally calculate the probability that the ($M,z$) distribution of observed clusters are not consistent with being drawn from the same parent population as the LP simulated clusters using the 2 dimensional Kolmogorov-Smirnov test (2d K-S)  \citep{1983MNRAS.202..615P,1987MNRAS.225..155F,2002MNRAS.335...73M}. The 2d K-S test is not sensitive to whether the total number of observed clusters is consistent with the total number expected, but is sensitive to comparisons of ($M,z$) distributions.

\subsection{Generation of Poisson simulated clusters}
\label{sim}\label{sssim}
The procedures described throughout the paper rely on the pixelisation of the mass-redshift ($M,z$) plane, and the calculation of the theoretical cluster abundance within each pixel, assuming particular survey geometries. 

For each pixelised map, we make at least one simulated distribution of clusters by Poisson sampling from the theoretically predicted cluster abundance value for each pixel and then rounded the value of the Poisson sample. We draw simulated clusters from those pixels whose Poisson sample are greater than zero. We randomly assign the final mass and redshift of the simulated cluster(s) such that they remained within the pixel boundaries. We pixelise the ($M,z$) plane within the region bounded by $7\times10^{13}<M_{200}<1.2\times10^{16}$ in mass and $1.0<z<2.2$ in redshift, with pixels of width $\Delta M=10^{11} M_{\odot}/h$ and $\Delta z=10^{-3}$ and used a 100 sq. deg. footprint to enable a direct  comparison with \cite{Jee}. We finally calculate $\R$ for each simulated cluster assuming best fit WMAP7 cosmological parameters.

We create $425$ such ($M,z$) distributions, by modifying the parameters of the Gaussian mass function by simultaneously Gaussian random sampling from the parameters $\Omega_M,\, \Omega_{\Lambda},\, \Omega_K\equiv(1-\Omega_M -\Omega_{\Lambda}), \, \Omega_b, \, H_0, \, \sigma_8, \, w_0, \, n_s,$ using the WMAP7 priors without imposing spatial flatness. For each set of standard cosmological parameters we additionally modify the amount of non-Gaussianity within the range $-200 < \fnl <400$ in steps of $\Delta\fnl=25$. 

We note that the total number of clusters obtained from the Poisson sampling of the mass function within the above survey geometry above $M>10^{14}\solM$, in the pixelised ($M,z$) plane $<386\pm107>$ agrees well with the theoretical expectation $<407\pm107>$, assuming the same input cosmology. We additionally create 350 sets of simulated clusters for use in \S\ref{suvgeom}, using the survey geometry $1.0<z<1.6$ and a footprint of 200 sq. deg.

%sim       386.74926+-       107.41557
%theory       407.46993+-       107.73848

\subsubsection{The Eddington bias}
The Eddington bias corrects for measurements made from non uniform population distributions.  The sharp decline of the cluster mass function with mass and redshift imply that more low mass clusters are likely to be scattered high, and masquerade as high mass clusters, than higher mass clusters are to be scattered low \citep[see e.g.,][]{2010arXiv1011.0004M}.   To mimic this bias in our simulations, we applied a mass error to each simulated cluster of the same magnitude as the average observed error ($41.8\%$), and then re-sampled the mass $M_{RS}$, and recalculated $\R$ for all clusters $M_{RS}>10^{14}\solM$. Below we refer to the simulated clusters obtained from this procedure as the re-sampled simulations. For comparison we present results of both ignoring, and including the Eddington bias. This shows how important this effect is, when comparing the observed Universe with simulations. We choose to apply the Eddington bias correction to the simulated clusters instead of the observed clusters.

\subsection{Comparing observations and simulations: ($>M,>z$)}
\begin{figure}
   \centering
\includegraphics[scale=0.4, clip=true, trim=20 15 15 35]{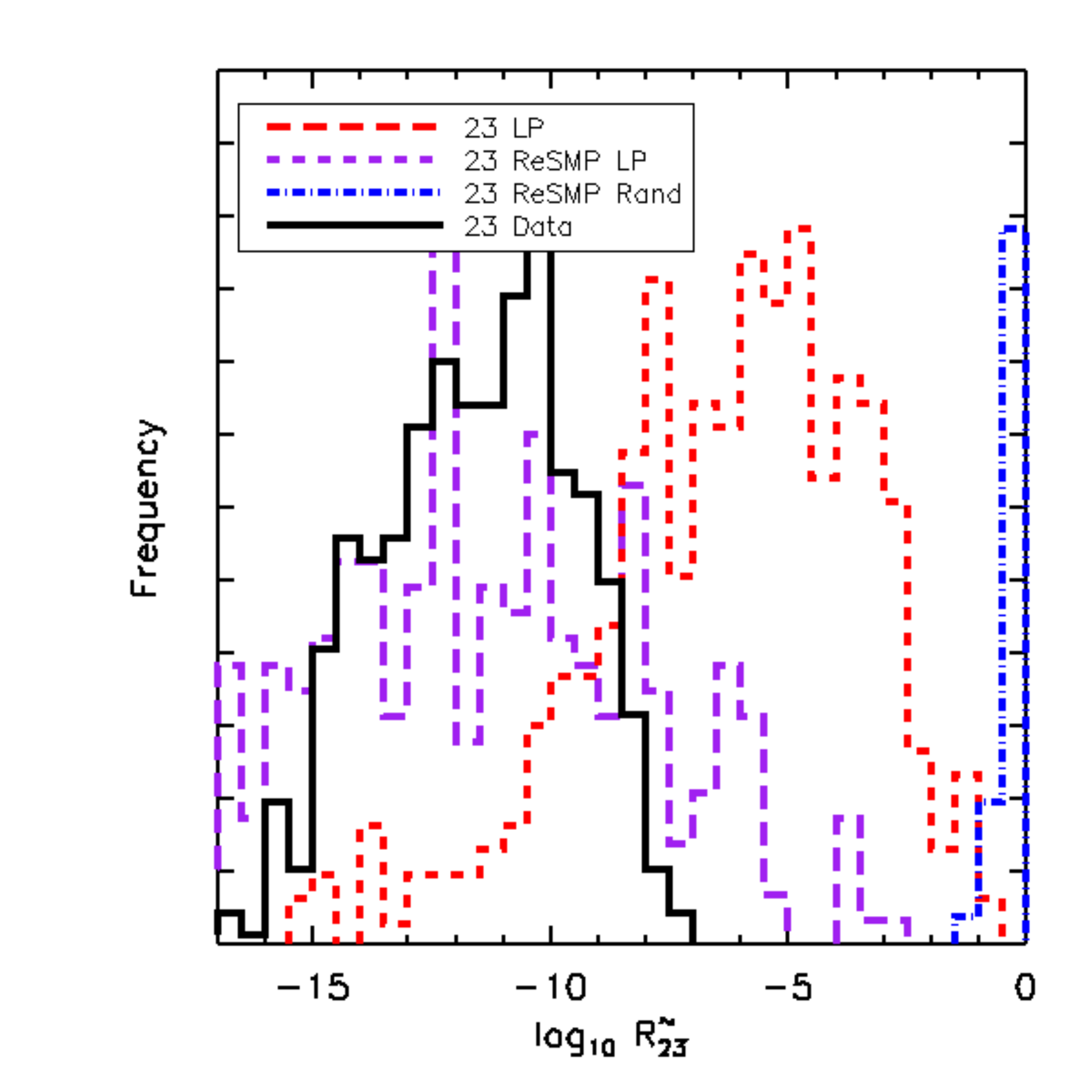}
   \caption{   \label{Pexist}  The distributions of $\R_{23}$ (in the $>M,>z$ sense) for ensembles of clusters. We show the distribution of the 23 LP clusters from each (re-sampled) simulation by the red (purple) dashed line, and the distribution of $\R_{23}$  for 23 randomly selected re-sampled clusters from each simulation by the blue dashed-dotted line. We show the values of $\R_{23}$ of the 23 observed clusters to exist by the black solid line. Using the 2d K-S test we ascertain that the observed clusters are inconsistent with the being the LP clusters, even though the values of  $\R_{23}$ agree.}
\end{figure}

If we multiply the ($>M,>z$) existence probabilities of the  clusters in Table 1 together, we obtain a range of values of $10^{-17}<\R_{23}<10^{-7}$ at 95\%, after marginalising over the cluster mass error. These values must be compared with $\R_{23}$ from simulated clusters before conclusions about tension can be drawn, see  \S\ref{mz}. In Fig. \ref{Pexist} we show the distribution of $\R_{23}$ for ensembles of simulated clusters. We show the distribution of $\R_{23}$ for the 23 LP (re-sampled) simulated clusters by the red dashed line (purple dashed line), and the distribution of  $\R_{23}$ for 23 random re-sampled  simulated clusters by the blue dashed-dotted line. The black solid line, corresponds to the values of $\R_{23}$ for the 23 observed clusters.

Examining Fig. \ref{Pexist} we conclude that if the observed clusters are compared with the LP simulated clusters, the $\R_{23}$ values are consistent with  $\Lambda$CDM at the 95$\%$ level, and comparing to the LP re-sampled simulated clusters (which includes the Eddington bias correction) we find the distributions of $\R_{23}$ to be completely consistent. Thus using rare clusters from H11 and \cite{Jee}, and the survey geometry specified in \cite{Jee}, which is less conservative than H11, we find no $\R_{23}$ tension of the observed clusters with $\Lambda$CDM. However if the clusters are actually drawn from a fair sampling of the mass function, we should instead compare the distribution of $\R_{23}$ with the randomly selected re-sampled simulated clusters (blue dash-dotted line). In this case we find the observed clusters are very inconsistent with the simulations. 

\subsubsection{The N$^{th}$ Least Probable clusters.}
\begin{figure*}
   \centering
\includegraphics[scale=0.35, clip=true, trim=20 15 15 35]{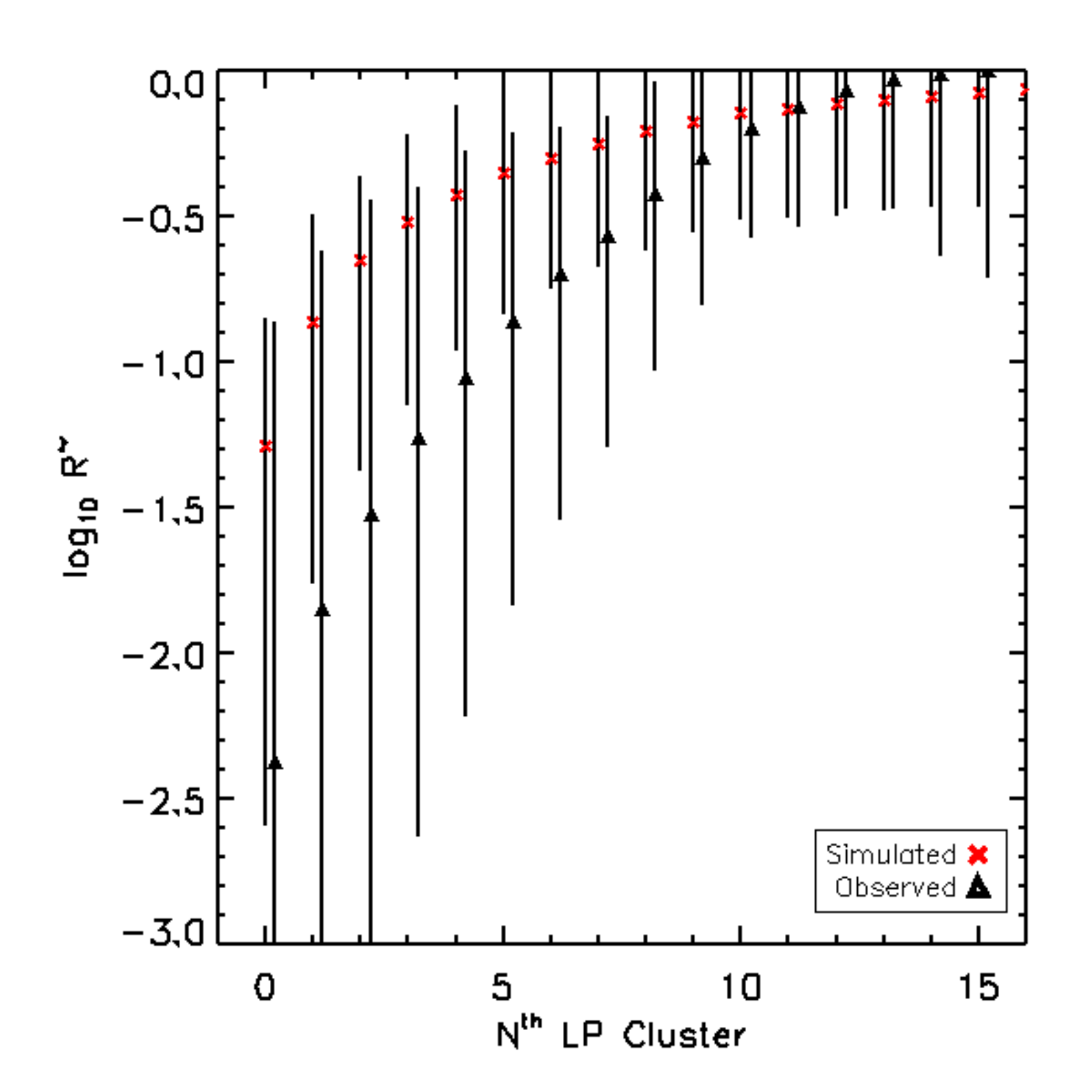}
\includegraphics[scale=0.35, clip=true, trim=20 15 15 35]{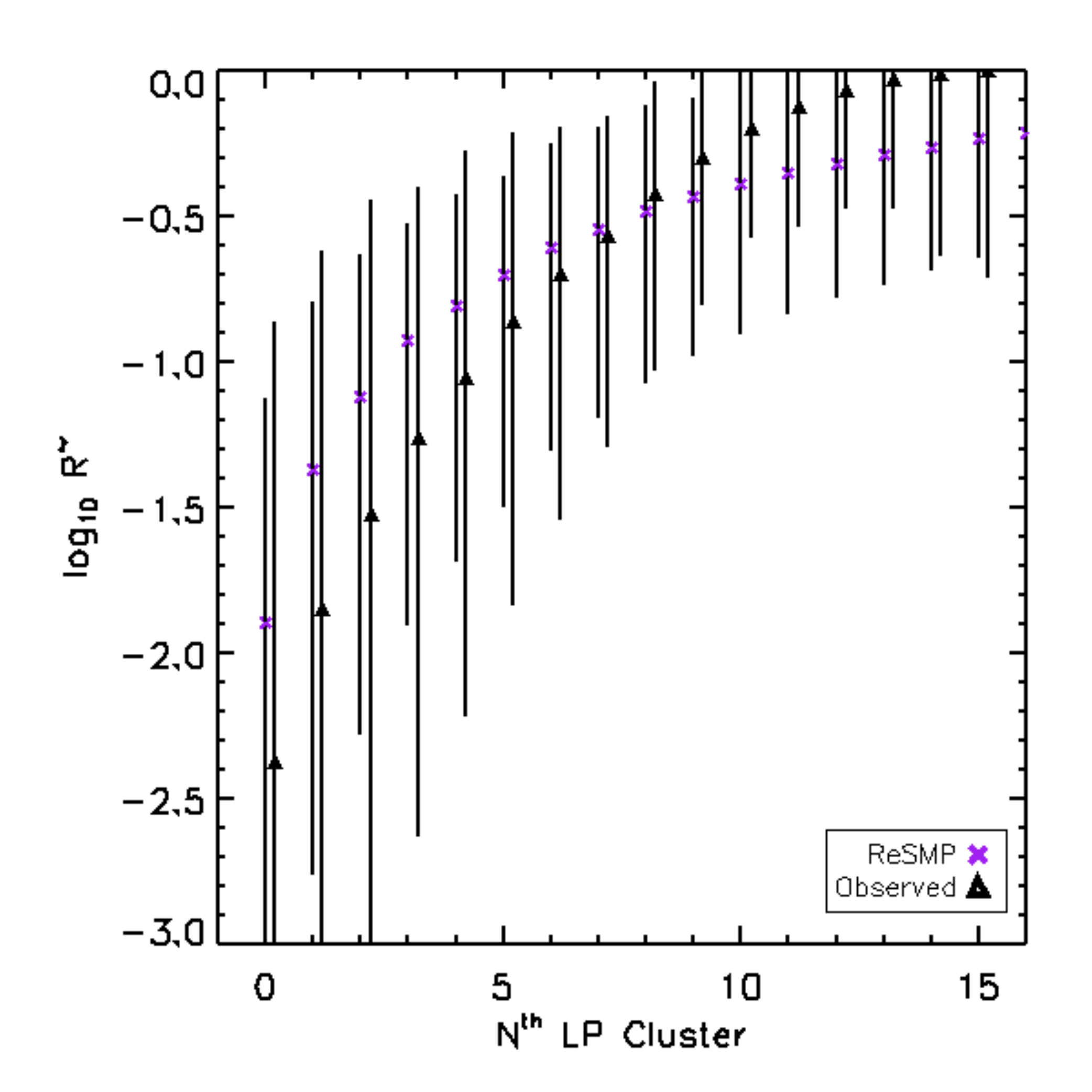}
   \caption{   \label{NLP22} The range of $\R$ values for each of the 17 least probable clusters drawn from the data and simulations (left panel) and the data and re-sampled simulations (right panel) after applying a $40\%$ mass error. The range of $\R$ values for the N$^{th}$ least probable cluster come from the variance between simulations after sampling from cosmological parameters, and from observations after sampling from the mass and mass error. The error bars show the spread of $\R$ at the 95\% confidence level.}
\end{figure*}

In Fig. \ref{NLP22} we show the range of  $\R$ values for each of the ranked N$^{th}$ least probable clusters from the simulations and the ranked re-sampled simulations, with the ranked values of $\R$ obtained by re-sampling from the mass and error of the observed clusters.

Fig. \ref{NLP22} again shows the importance on the distribution of $\R$ values of comparing the observations with the re-sampled simulations, which account for the Eddington bias.  Examining the right panel of Fig. \ref{NLP22} we see that the first 8 clusters all have lower $\R$ values than the 8 least probable re-sampled simulated clusters. One may choose to only analyse and compare these 8 observed clusters, which have the lowest $\R$ values,  with the 8th least probable simulated clusters.  This will remove any weakening of the true level of tension these clusters may present, by including clusters which are common (and thus have high $\R$ values). 

\begin{figure} 
  \centering
\includegraphics[scale=0.4, clip=true, trim=20 15 15 35]{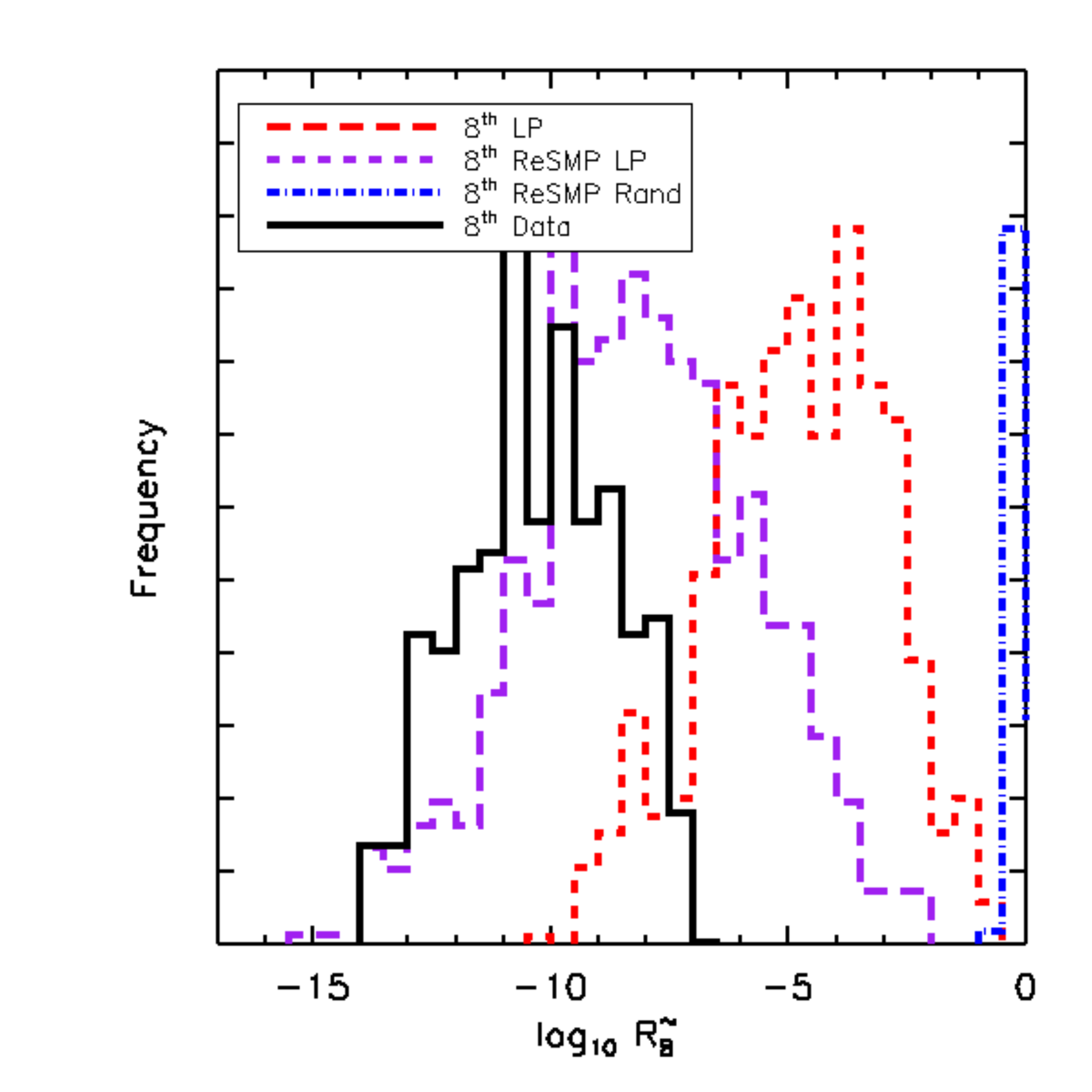}
   \caption{   \label{Pexist2}  The distributions of $\R_{8}$ (in the $>M,>z$ sense) for ensembles of clusters. We show the distribution of the 8 LP clusters from each (re-sampled) simulation by the red (purple) dashed line, and the distribution of $\R_{8}$  for 8 randomly selected re-sampled clusters from each simulation by the blue dashed-dotted line. We show the values of $\R_{8}$ of the 8 observed clusters by the black solid line. }
\end{figure}

In Fig. \ref{Pexist2} we show the $\R_8$ distributions of the 8 observed clusters with the lowest $\R$ values, with the 8th LP simulated (and re-simulated) clusters, and 8 randomly selected re-sampled simulated clusters. Even after excluding clusters which are more common (with large $\R$ values), we continue to find that the observed clusters have $\R_8$ values in agreement with $\Lambda$CDM at the $1\sigma$ level by comparing with the LP re-sampled simulated clusters.  

We remind the reader that in performing this analysis, we have fully marginalised over the observed cluster mass error, the simulated cluster mass error, and the cluster mass function assuming WMAP7 cosmological priors without imposing spatial flatness.  We cannot substantiate the claim that the observed clusters are the LP clusters, without further analyses, or without complete follow up of all clusters in the footprint. We therefore suggest alternative tests to determine if the observed clusters are consistent with being drawn from the same parent population as the LP clusters.

\subsection{Comparing observations and simulations: redshift histograms}
\begin{figure}
   \centering
   \includegraphics[scale=0.4, clip=true, trim=20 15 15 35]{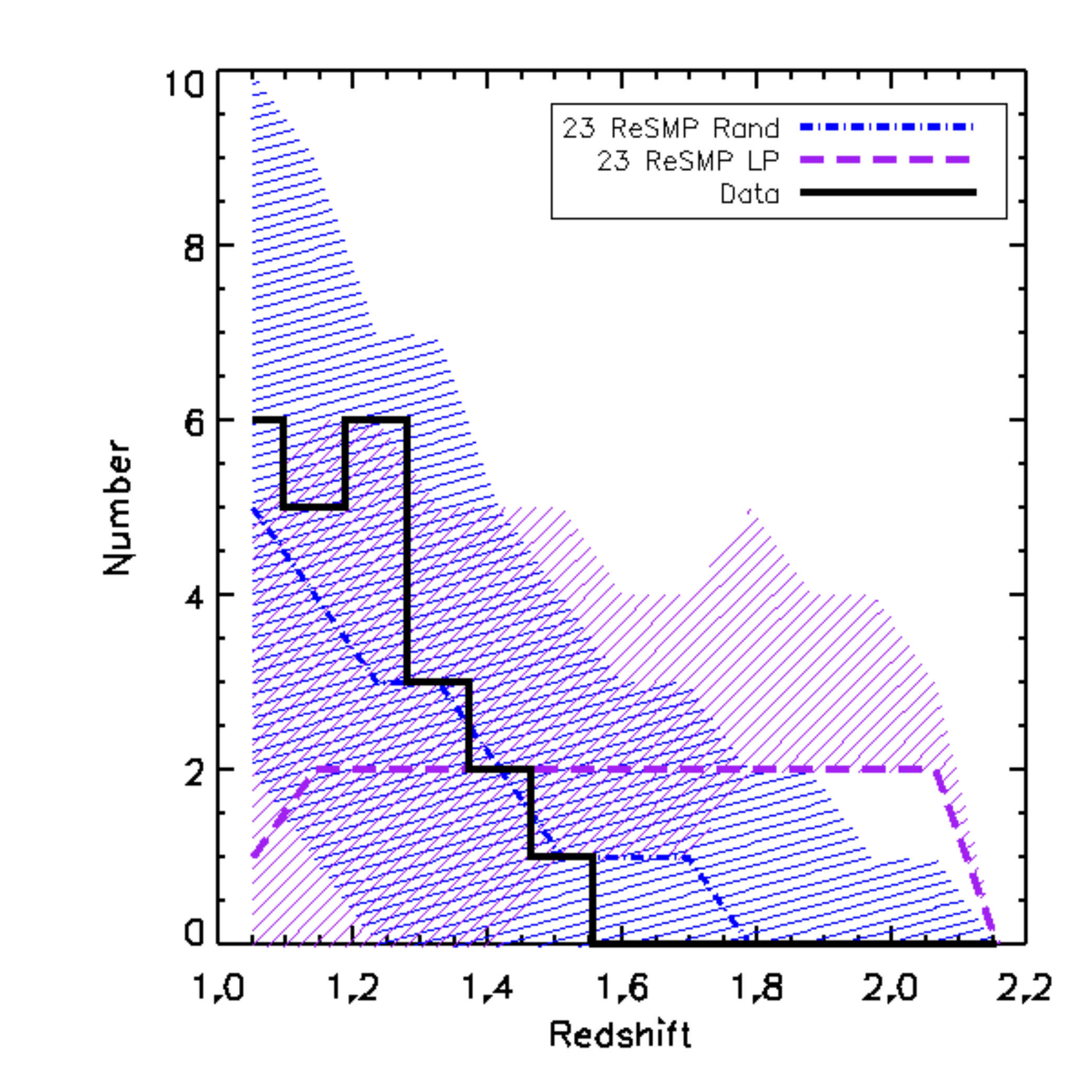}
   \caption{   \label{RedDist} The redshift distribution for the 23 observed clusters (black solid lines), 23 randomly selected re-sampled clusters (blue dot-dashed line), and 23 LP re-sampled clusters (purple dashed line). The lines show the median of the distributions and the shaded regions show the 95\% spread of the distribution.}
\end{figure}

In Fig. \ref{RedDist} we show the redshift distribution of the 23 observed clusters (black solid lines) and 23 randomly selected re-sampled clusters (blue dot-dashed lines) and 23 LP re-sampled clusters (purple dashed line). The lines show the median of the distributions and the shaded regions show the 95\% spread of the distribution. The simulations were performed assuming the redshift range $1<z<2.2$.  

We find that the redshift distributions of the re-sampled LP clusters is inconsistent with the observed data at the $>$95\% confidence level. If the true survey geometry is $1<z<2.2$ and the clusters are consistent with being the LP clusters, we would expect $\gtrsim 8$ of the 23 observed clusters to have a redshift greater than $1.6$ and we observed none, resulting in Poisson expectation probability $\Po(0,8)=\exp(-8)$.  By examining the distributions of randomly selected re-sampled clusters with the observed data, we find that these distributions are in agreement.  

$\,$\\
This leads us to an interesting conclusion: If the survey geometry is $1<z<2.2$  and the observed clusters are compared with being the (Eddington bias corrected) re-sampled LP clusters, then the values of $\R_{23}$ are in agreement with those predicted from a $\Lambda$CDM model. However, the observed redshift distributions and the $\Lambda$CDM-predicted redshift distributions are not in agreement at the level of $\sim\exp(-8)$. If, however, the observed clusters are compared to a random selection of the re-sampled simulated cluster population, then the values of $\R_{23}$ are highly inconsistent with those predicted by $\Lambda$CDM-predicted at $\gtrsim 3 \sigma$, but the redshift distributions agree.

We further compare the ($M,z$) distributions of observed and re-sampled simulated clusters using the 2d K-S test.

\subsection{Comparing observations and simulations: 2d K-S test}
\label{2dkstestddes}

Different distributions in the 2 dimensional ($M,z$) plane can yield the same product of existence probabilities $\R_{23}$, but the full distribution in the ($M,z$) plane carries much more information than just comparing $\R_{23}$.  The  2d K-S test is well suited to extract this information \cite{2002ApJ...581....5V}.

To determine if the distribution of observed clusters is consistent with being the LP re-sampled clusters from simulations, or a random selection of re-sampled simulated clusters, we compared their ($M,z$) distributions using the 2d K-S test. 

The null hypothesis of the 2d K-S test asserts that two 2 dimensional data samples are drawn from the same parent distribution. Two distributions are said to have been drawn from the same parent population if the 2d K-S test probability is $>0.2$ ($>10^{-0.7}$).

We compared the ($M,z$) distributions of 23 LP re-sampled clusters between simulations (Sim P$_{\mathrm{LP}}$), and with $100$ realisations of the observed  clusters D$^{\mathrm{x}}$. Each realisation sampled from the cluster mass and error to produce a sampled mass. 
Finally we compared two randomly selected sets of re-sampled clusters with each other (Sim P$_{\mathrm{RAND}}$) and with the observed clusters. We present the resulting 2d K-S test probabilities in Table \ref{2dksprobtable}.
  
\begin{table}
   \centering
  \begin{tabular}{|  c  |  c |  c | c |} 
  \hline
S1(M,z) & S2(M,z) &  $<\log$P$>$ $\fnl^{-200}$  &  $<\log$P$>$ $\fnl^{0}$  \\ \hline
Sim P$_{\mathrm{LP}}$ & Sim P$_{\mathrm{LP}}$  & $-0.79 \pm 0.67$& $ -0.81 \pm 0.72$ \\
D$^{\mathrm{x}}$ & Sim P$_{\mathrm{LP}}$  & $-3.24\pm0.97$& $ -3.33\pm0.96$ \\
 D$^{\mathrm{x}}$ & Sim P$_{\mathrm{RAND}}$    & $-5.09\pm1.08$& $-4.94\pm1.08$ \\ \hline
S1(M,z) & S2(M,z) &  $<\log$P$>$ $\fnl^{200}$  &  $<\log$P$>$ $\fnl^{400}$  \\ \hline
Sim P$_{\mathrm{LP}}$ & Sim P$_{\mathrm{LP}}$  &  $-0.82 \pm 0.70$ & $-0.84 \pm 0.73$ \\
D$^{\mathrm{x}}$ & Sim P$_{\mathrm{LP}}$  &  $-3.36\pm0.94$& $-3.50\pm0.91$\\
 D$^{\mathrm{x}}$ & Sim P$_{\mathrm{RAND}}$    &  $-4.85\pm1.186$& $-4.70\pm1.13$ \\ \hline
  \end{tabular}
    \caption{  \label{2dksprobtable} We show the average probability that the 2 dimensional ($M,z$) distribution of clusters in sample 1 (S1)  and sample 2 (S2)  is consistent with being drawn from the same parent population, using the 2d K-S test. Each of the 425 simulations were performed by sampling from the cosmological parameters with WMAP7 priors.   D$^{\mathrm{x}}$ denotes the observed data and the P subscript indicates how the re-sampled simulated clusters (of equal number to the observed clusters) were selected e.g., the least probable (LP) or randomly (RAND).}
\end{table}

The average  2d K-S test probability that both sets of LP re-sampled simulated clusters is drawn from the same parent population is $\gtrsim 0.2$, which implies consistency with being drawn from the same population. This was of course to be expected, as each simulated set of LP clusters were drawn from a Poisson sampling of the mass function with only small changes to the cosmological model parameters within the WMAP7 priors. The probability that the observed clusters and the LP re-sampled simulated clusters were drawn from the same parent population is $\sim 0.003$, i.e., the observed clusters are not consistent with being the LP clusters.

Returning to Table \ref{2dksprobtable}, if we examine the 2d K-S test probability obtained by comparing the LP and observed clusters for increasing values of $\fnl$, we see the probability slowly decreasing, i.e. the clusters are becoming less consistent with the simulated LP simulations as $\fnl$ increases. This can be understood if the observed clusters are lower in mass and redshift than the LP clusters, because increasing $\fnl$, moves the set of LP clusters to higher masses and redshifts.

We next compared the distributions of sets of randomly selected simulated clusters with one another. As expected, we again found the probability to be consistent with being drawn from the same parent distribution. The comparison between randomly selected simulated clusters with the observed selected clusters produced a probability lower than that of the simulated LP clusters. This suggests that while the observed clusters are inconsistent with being the LP clusters, they are skewed towards the LP clusters from a purely random distribution.

If one has a complete sample, one may of course perform the full 2d K-S test comparing the distributions of observed clusters with those obtained from simulations. Furthermore, if we incorrectly assume the shape of the survey geometry our results may be also biased. We explore this effect further below, using a sample of only X-ray selected cluster, and in a future paper (Hoyle et al. in prep.).

\subsection{An analysis with X-ray selected clusters.}
\label{suvgeom}
\begin{figure*}
   \centering
   \includegraphics[scale=0.4, clip=true, trim=35 20 25 35]{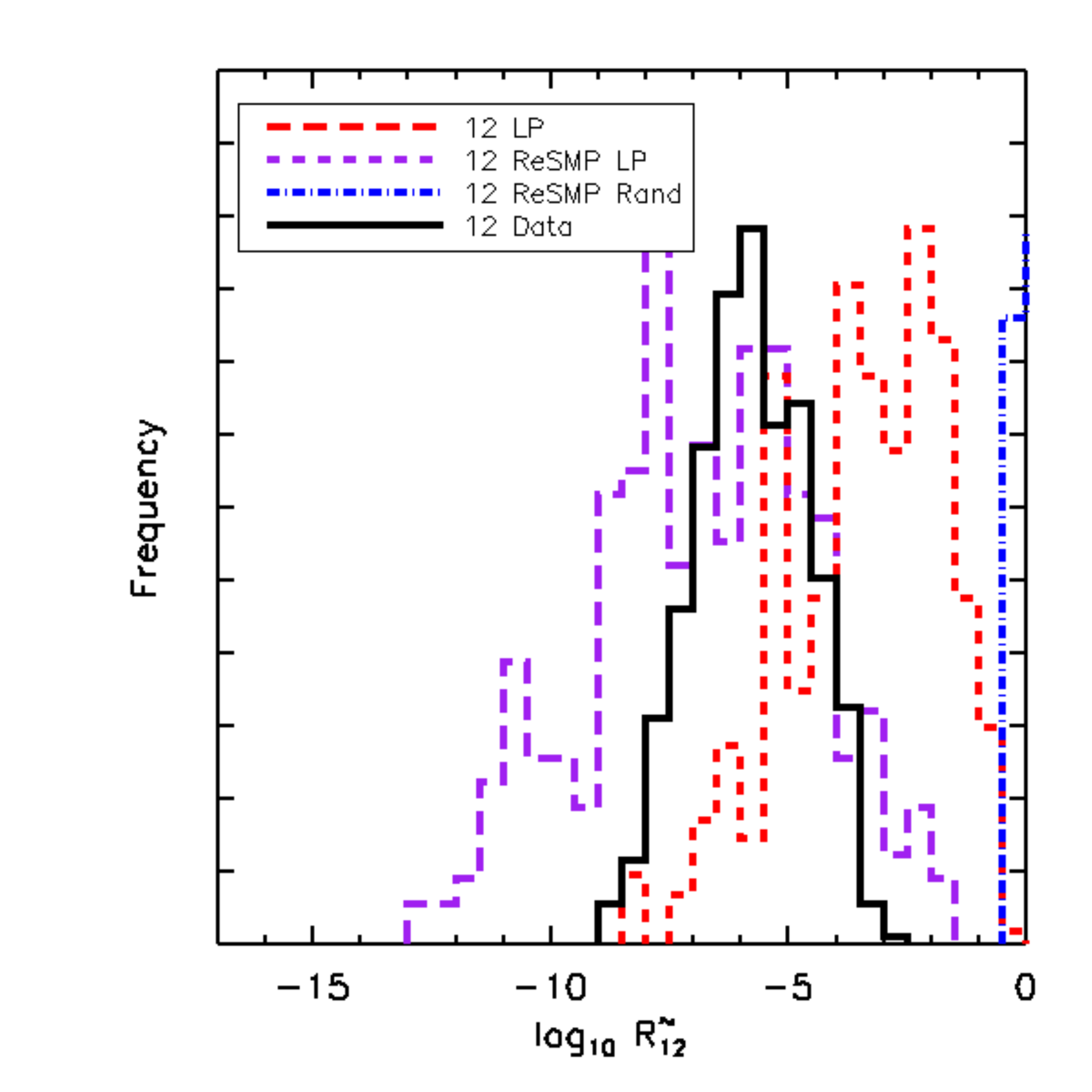}
\includegraphics[scale=0.4, clip=true, trim=35 20 25 35]{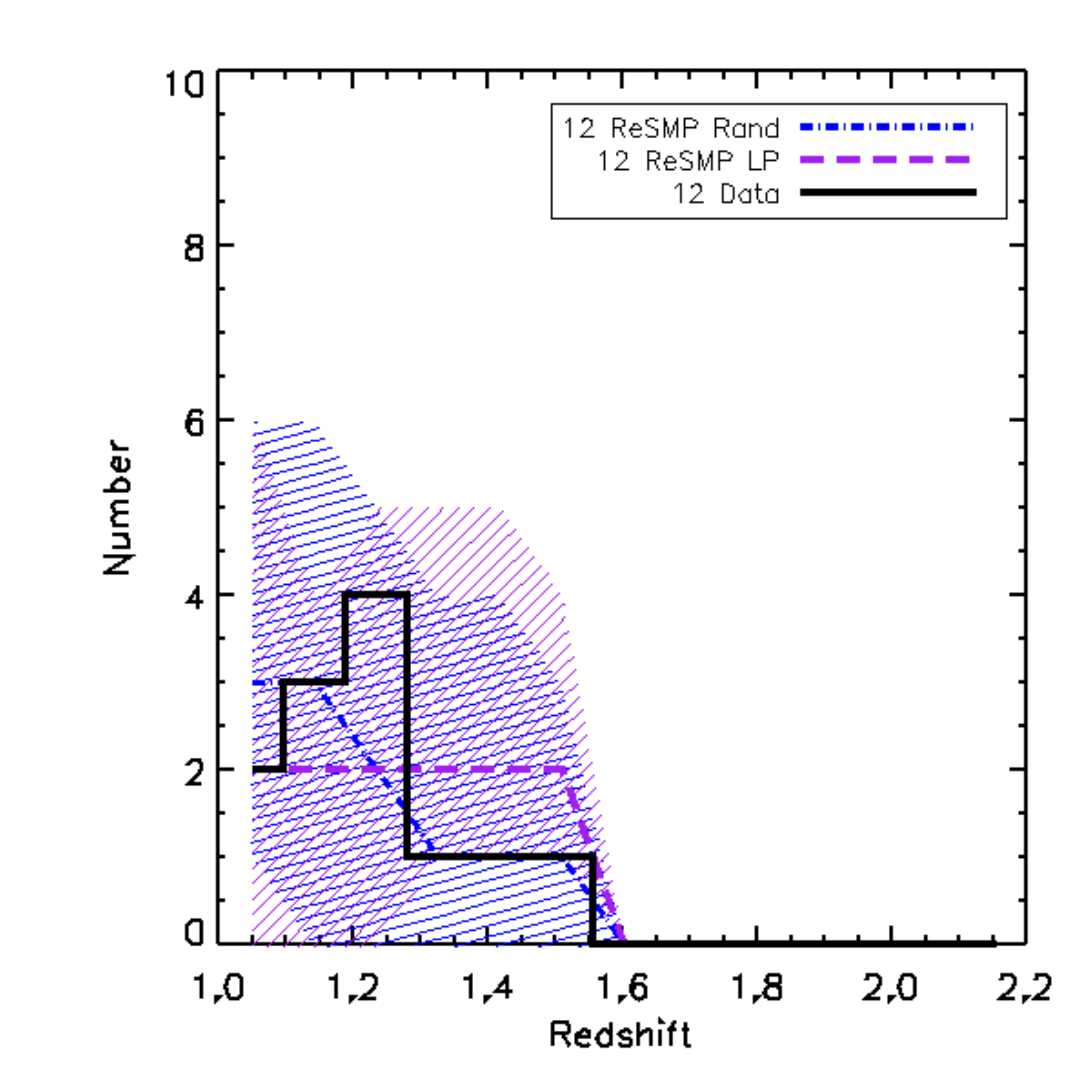}
   \caption{   \label{UpdGeom} Analysis using the 12 X-ray selected clusters. Right panel: As Fig. \ref{Pexist}. Left panel: As Fig. \ref{RedDist}. Both panels are compared with simulations using  a survey geometry with a redshift range of $1.0<z<1.6$ and a footprint of 200 sq. deg.}
\end{figure*}

We next compare the sub sample of 12  X-ray selected clusters from Table \ref{highzclustable}, with 300 sets of simulated clusters assuming a survey geometry with a redshift range of $1<z<1.6$ with an increased survey footprint of 200 sq. deg using the statistical tests introduced above. In Fig.\ref{UpdGeom} we show the results of the $\R_{12}$ existence probability distributions, and the redshift histogram analyses. 
We find that the distributions of $\R_{12}$ for the observed and re-sampled LP simulated clusters (right panel of  Fig.\ref{UpdGeom}) are consistent at the $<$95\% confidence level. The left panel of  Fig.\ref{UpdGeom} shows that the redshift distributions are in agreement at the $<$95\% confidence  level with the LP (and random) simulated clusters. Furthermore, the 2d K-S test probability that the 12 X-ray selected clusters are drawn from same parent population as the 12 LP re-sampled simulated clusters, within this new survey geometry, is $10^{-0.91\pm 0.46}$ (the quoted error is $1\sigma$), which shows consistency at the $<95\%$ confidence level. Recall consistent 2d K-S test probabilities are $\sim10^{-0.7}$. The 2d K-S test probability that the X-ray selected clusters are consistent with being drawn from the same parent population as the randomly selected re-sampled simulated clusters is $10^{-2.87\pm 0.72}$, which is inconsistent at the $>95\%$ confidence level.

While this redshift cut, and larger survey footprint does remove tension (to  $<$95\%) on all statistics when comparing the X-ray selected observed clusters with the re-sampled simulated LP clusters, we do however note that higher redshift clusters have been found in surveys which also have X-ray emission, and could conceivably be found in X-ray selected surveys, e.g.,  CL J1449+0856 is a cluster of mass $5-8\times10^{13}\solM$ at $z=2.07$ \cite{2011A&A...526A.133G}.

\section{Conclusions and discussion}
\label{conclusions}
The observations of massive, high redshift clusters and their interpretation within the standard cosmological framework is currently a hot topic in observational cosmology. Current and future surveys will (probably) identify more of these massive clusters, and a robust framework is required to describe the level of agreement or tension these clusters cause with our current model of cosmology. Recently \citep[][hereafter H11)]{2010arXiv1009.3884H}, \citep{2010arXiv1012.2732E,Jee} used high redshift massive clusters to indentify large levels of tension with the $\Lambda$CDM model assuming WMAP priors on cosmological parameters.

We present a critical re-evaluation of the current statistical techniques used as a means to measure this tension with $\Lambda$CDM, and show their range of validity, paying particular attention to the ($>M,\,>z$) analysis and related exclusion curves \citep{2010arXiv1011.0004M}. 

We show that the ($>M,>z$) analysis, although easily calculated, is not equivalent to the level of tension present with a model. This bias was first addressed in \cite{2011arXiv1105.3630H}. We demonstrate that both the ($>M,>z$) analyses and related exclusion curves, can be used to measure tension with a model, but only after carefully calibrating with simulations. In particular we critically analyse seemingly rare clusters from H11 and \cite{Jee} using the survey geometry specified in \cite{Jee} of $1.0<z<2.2$ with a footprint of 100 sq. deg., which is less conservative than H11 due to the smaller footprint.

We compile a list of high redshift ($z>1.0$), massive ($M>10^{14}\solM$) galaxy clusters from the literature. Most (20/23) of the clusters have weak lensing mass measurements from \cite{Jee}.  We Poisson sampled from the theoretical cluster mass function, assuming the above survey geometry, to create $425$ sets of simulations, each for a range of values of primordial non-Gaussianity described by  $-200<\fnl<400$. The cosmological parameters ($\Omega_{\Lambda}, \,\Omega_m, \, h, \, n_s,\, \sigma_8$) for each simulation are drawn randomly from the WMAP7 priors without assuming spatial flatness. We address the Eddington bias by allocating a mass error to each of the simulated clusters, and re-sampling their masses, and refer to these sets of simulations below as the re-sampled simulations. The magnitude of the simulated cluster mass error is chosen to be the averaged observed mass error.

We compare the distributions of existence probabilities $\R$, and ensemble existence probabilities $\R_{23}$, calculated in the  ($>M,>z$)  sense, between the 23 observed clusters, with sets of the 23 Least Probable (LP) clusters from each simulation, and with sets of 23 randomly selected clusters from each simulation. We found that if (and only if) the clusters were compared with the re-sampled LP clusters their values of $\R_{23}$ present no tension with $\Lambda$CDM with WMAP7 priors on cosmological parameters, but if they were drawn from a purely random sample of clusters they presented a discrepancy of $\sim 2 \sigma$, depending exactly on the unknown selection function. We further analyse only the observed clusters with the lowest $\R$ values, and find that their $\R$ values still present no tension with the re-sampled LP clusters. 

We next analyse the redshift distributions of the observed and least probable re-sampled simulated clusters, and find them to be inconsistent at the level of $\sim\exp(-8)$. The redshift distributions of the observed clusters are consistent with the randomly selected clusters at $<95\%$ confidence.

This leads us to an interesting conclusion: If the survey geometry is $1<z<2.2$  and the observed clusters are compared with being the (Eddington bias corrected) re-sampled LP clusters, then the values of $\R_{23}$ are in agreement with those predicted from a $\Lambda$CDM model. However, the observed redshift distributions and the $\Lambda$CDM-predicted redshift distributions are not in agreement at the level of $\sim\exp(-8)$. If, however, the observed clusters are compared to a random selection of the re-sampled simulated cluster population, then the values of $\R_{23}$ are highly inconsistent with those predicted by $\Lambda$CDM-predicted at $\gtrsim 3 \sigma$, but the redshift distributions agree.

We next show that the ($>M,>z$) $\R$ statistic encodes mass and redshift information into one number. However the clusters produce allowable values of $\R$, but appear to have a strange redshift distribution.  The full 2d ($M,z$) distribution of clusters is to be more carefully compared with simulations using the 2d K-S test, which is sensitive to inconsistent distributions.

Using the 2 dimensional Kolmogorov-Smirnov (2d K-S) test, we formally calculate the probability that the ($M,z$) distribution of observed clusters is consistent with being drawn from the same parent population as the distribution of re-sampled simulated LP clusters, or from sets of randomly-selected re-sampled simulated clusters. We found that the observed clusters are inconsistent $\Po=10^{-2.47 \pm 0.90}$ with being drawn from the LP clusters, but are likely to be skewed towards the LP clusters from the random sample of clusters which have a lower probability of $\Po=10^{-4.94 \pm 1.08}$.

To summarise our main results, if we assume the survey geometry presented  \cite{Jee},  we find that if the observed clusters were compared to the re-sampled LP simulated clusters, the $\R$ statistics are consistent, but the redshift distributions and the 2d K-S probabilities were highly inconsistent. If, however, we compared the observed clusters to randomly selected re-sampled simulated clusters, the redshift distributions are consistent, but the $\R$ statistic and the 2d K-S test probability are highly inconsistent. Using the 2d K-S test we show that this problem cannot be solved by including non-Gaussianity within the range of $-200<\fnl<400$, and may be due to the non trivial selection function and survey geometry of the heterogeneous cluster sample. 

We finally examine a sub sample of the 12 X-ray selected clusters with simulated clusters drawn from simulations performed assuming a more conservative survey geometry described by a redshift range of $1<z<1.6$ and a footprint of 200 sq. deg. The modified survey geometry may account for the difficulty of obtaining galaxy redshifts at higher redshift, and differences in (overlapping) X-ray survey footprints and flux limits. We find that with this more conservative analysis, no tension is present if the observed X-ray selected clusters are compared with the LP re-sampled simulated clusters, and have consistent values of $\R_{12}$, redshift distributions, and 2d K-S test probabilities. That high-z X-ray selected clustesr are consistent with being the LP clusters, may be indicative of a publication bias, i.e., only the most interesting (or massive) systems are being published. We examine the possibility of obtaining constraints on model parameters from cluster surveys with an unknown selection function using {\it a posterior} statistics, in a follow up paper (Hoyle et al. in prep.).

\section*{Acknowledgements} 
\label{ack}
The authors would like to thank an anonymous referee for useful comments and suggestions. BH would like to thank Christian Wagner and Jorge
Norena for detailed discussions and Christian Wagner for making the results of his simulations available. BH acknowledges grant number FP7-PEOPLE-2007- 4-3-IRG n 20218, and MICINN grant AYA2008-0353, and would like to thank the mathematics department at the University of Cape Town for hospitality. LV and RJ are supported by MICINN grant AYA2008-0353. LV is supported by
FP7-IDEAS-Phys.LSS 240117. SH is supported by the Academy of Finland grant 131454.
%BIBLIOGRAPHY
\bibliographystyle{JHEP}

\bibliography{hz2}

%\section{Combining probabilities}
%\label{flip}
%To test the validity that multiplying the ``probability of existence'' of individual cluster to obtain an ensemble of existence, we simulated a long chain of coin tosses and record the outcome, but note this works equally well with a chain of dice rolls or systems with an arbitrary number of outcomes. We intentionally only examine a finite section of the simulated chain  (e.g., from a starting position $p$, to the end of the chain) which can be thought of as corresponding to the redshift slice of interest. We examine the chain for sub-chains of a repeated number (of length at least $n$) of repeated outcomes O$^{>n}_{>p}$, in our example, if n$=4$ and $p=0$, we identify all such sub-chains like HHHH or TTTTTT etc., along the length of the chain. The final probability of O$^{>n}_{>p}$ is simply the number of such events in any chain divided by the number of simulated chains. 
%We then repeat the analysis from a new starting position $p1$, and identify any $>n1$ repeated outcomes  O$^{>n1}_{>p1}$. We repeat this process $m$ times at random starting positions, looking for randomly assigned $>n_m$ repeated outcomes   O$^{>n_m}_{>p_m}$.

%Table \ref{coinflip} shows the probability that any individual set of O$^{>n}_{>p}$ system exists when identified alone, and also the inclusive [{\bf check terminology}] probability $P^{\&}$ that all O$^{>n0}_{>p0} \,\&$  O$^{>n1}_{>p1}\,\& $ ... O$^{>n_m}_{>p_m}$ along any chain, and the percentage difference between niavely multiplying the probabilities together, and the inclusive probability.

\end{document}